\documentclass[onecolumn,superscriptaddress,amsmath,amssymb,floatfix]{revtex4}

\usepackage{graphicx}
\usepackage{amssymb}
\usepackage{amsmath}
\usepackage{epstopdf}
\usepackage{mciteplus}
\usepackage{setspace}
\usepackage{textcomp}
\usepackage{setspace}

\begin{document}
\title{Validating pore-scale models of drying using microfluidic experiments}

\author{Paolo Fantinel}
\affiliation{Max Planck Institute for Dynamics and Self-Organization(MPIDS), 37077 G\"ottingen, Germany}

\author{Oshri Borgman}
\affiliation{Department of Soil and Water Sciences, The Hebrew University of Jerusalem, Israel}

\author{Ran Holtzman}
\affiliation{Department of Soil and Water Sciences, The Hebrew University of Jerusalem, Israel}

\author{Lucas Goehring}\email{lucas.goehring@ntu.ac.uk}

\affiliation{Max Planck Institute for Dynamics and Self-Organization(MPIDS), 37077 G\"ottingen, Germany}
\affiliation{School of Science and Technology, Nottingham Trent University, Clifton Lane, Nottingham, NG11 8NS, UK}

\begin{abstract}
We present an experimental micro-model of drying porous media, based on microfluidic cells made of arrays of pillars on a regular grid, and complement these experiments with a matching two-dimensional pore-network model of drying.  Disorder, or small-scale heterogeneity, was introduced into the cells by randomly varying the radii of the pillars, around their average value.  The microfluidic chips were filled with a volatile oil and then dried horizontally, such that gravitational effects were excluded. The experimental and simulated drying rates and drying patterns were then compared in detail, for various levels of disorder, in order to verify the predictive capabilities of our model. The geometrical features were reproduced well, while reproducing drying rates proved to be more challenging. 
\end{abstract}

\maketitle

\section{Introduction}
 
Drying of porous materials is an important example of a two-phase flow process that occurs in both natural and engineered systems \cite{Lehmann2008}; it is a fundamental process in soil-atmosphere energy and moisture exchange \cite{MeteorologyCRPFRP}, in solute transport within soils \cite{nachshon2011combined} and in agriculture \cite{bernstein1975effects}. In such situations drying can be approached as an immiscible displacement problem where an invading fluid (air) penetrates the medium by displacing a more viscous defending fluid, as the latter evaporates \cite{Shaw}. During drying, periods of evaporation from a static air-liquid interface are interrupted by sudden invasion events called Haines jumps \cite{Haines1930,bursting}. When such jumps, or bursts, occur, pore-scale interfaces attached to the grains depin and advance to their next stable position. This motion induces liquid flows that can cause connected interfaces elsewhere to readjust, often on a very short timescale~\cite{bursting,Armstrong2013}. 

Drying in porous materials can be separated into two different stages, based on their characteristic drying rates and transport mechanisms. During the first stage liquid is transported mostly via flow from the bulk, through connected liquid pathways, to the medium's surface where evaporation occurs \cite{Shokri2010a}. During this time the evaporation rate is fairly constant (constant rate period) and influenced mostly by surface wetness, the size distribution of surface pores and the surface boundary conditions, e.g. wind \cite{Shahraeeni,Suzuki1968,lehmann2013effect}. Stage two, or the falling rate period, is identified with the loss of connectivity between the exposed surface and the liquid in the pore-space. As the fluid interface recedes into the porous medium, relatively slow vapor diffusion becomes the dominant transport mechanism \cite{Chauvet2009,shokri2011determines} causing the drying rate to drop noticeably.

Experimentally, progress in understanding the basic physics of transport in porous media has been made in different ways including the use of etched channel networks \cite{ExpVorhauer,Laurindo1998,vorhauer2015drying}, Hele-Shaw cells containing cylindrical pillars \cite{Ferrariexp,aursjo2011direct} and monolayers of silica spheres \cite{Shaw,expglassbeads}. These experiments have typically focused on long throat networks \cite{ExpVorhauer,Laurindo1998,vorhauer2015drying}, scales that are significantly larger than real soils \cite{Ferrariexp} or pore networks with only a few features \cite{12experiment}, due to the challenges of solving manufacturing problems that arise when precisely crafting many micron-sized objects.  

Computationally, modeling a three dimensional porous system is also still a difficult task, although many attempts have been made in this direction (see e.g. \cite{bursting}). This is due to the need to limit numerical resolution when dealing with such a high number of variables \cite{Ferrariexp}. A possible solution is to approach the problem from a two-dimensional description and, once the main features are captured, later expand the model into a three-dimensional approach. Pore-network models~(PNM), originally proposed by Fatt et al.~\cite{fatt1956network}, have become accepted in modeling multi-phase flow in porous media \cite{blunt2002detailed}, thanks to their combination of computational efficiency and the ability to capture the essential pore-scale physics, while coarse graining over sub-pore effects. Early applications to drying were made by Nowicki et al. \cite{nowicki1992microscopic} and Prat \cite{Prat1993}. They were able to calculate effective permeabilities \cite{nowicki1992microscopic} and to estimate the stabilizing effects of gravity on the invasion front into a drying body \cite{Prat1993}. Generally, PNM  of drying arrays of pores and throats are based on mass balances of liquid and vapor in single pores: transport can occur through diffusion of vapor and viscous liquid flow (e.g.~\cite{Yiotis2001}). PNM have also shown the effects of capillary pumping on invasion patterns \cite{tsimpanogiannis1999scaling} and how liquid flow through films in the corners of angular pores can modify drying rates \cite{Yiotis2004}. Furthermore, in \cite{Laurindo1998} a model is proposed where the porous medium is coupled with an external, diffusive, boundary layer, allowing for solutions of complex boundary layer conditions and processes. The strength of PNM lies in their ability to capture complex effects very efficiently, allowing one to investigate the minimum amount of assumptions needed to accurately describe, or predict, observational results.

In this paper we aim to connect pore-scale observations of drying phenomena to their macroscopic interpretation by introducing a type of experimental micro-model that is based on microfluidic techniques. We also use these micro-models to test and validate a complimentary PNM. The novelty in this method is the ability to manufacture large two-dimensional porous systems with complete control of features at a scale of a few micrometers, a length-scale comparable to those often directly relevant in geological and agricultural applications. Furthermore, we will show that unprecedented control over the size and position of the pores is now achievable and how the possibility to reproduce the experimental geometries exactly in the simulations allows for direct comparisons between experimental and simulated results. 

 We first explain both the experimental and simulation procedures. We then show our results and, finally, discuss them in the last section of this paper. Finally, we note that the complementary PNM will be detailed and explored further in a companion paper \cite{Borgman2016}, which will also deal with the effects of local correlations in disorder on drying.

\section{Methods}
 
\subsection{Microfabrication}
Pseudo-2D micro-mechanical models were produced with standard microfluidic techniques including soft-lithography. For a review of these methods, see Ref.~\cite{Madou}. We start by spin-coating a negative photoresist (SU8 3025, MicroChem Corp.) onto a silicon wafer, to obtain a flat layer of nominal thickness $h = $ 40 $\mu$m. The resist is then exposed to UV light through a mask reproducing the desired pattern. The areas exposed to light crosslink and the unexposed areas are removed by rinsing the sample with a developing solvent (mr-Dev 600, micro resist technology GmbH) and isopropanol, leaving the desired pattern of SU8 on the wafer. We use this raised SU8 structure as a mold, or master, onto which we pour liquid poly(dimethylsiloxane), PDMS, an elastomer that is then cured for one hour in an oven at 75\textdegree C. We peel the cured PDMS layer from the master and use it as a further mold for a layer of Norland Optical Adhesive 81 (NOA, Sigma-Aldrich)\cite{NOA63,NOA81}, which will form the body of one of our microfluidic chips.  A base for the chip is then prepared by coating an acetate sheet with NOA. We initiate the curing of the two parts of our chip by exposing them to UV light for about 10 seconds. Once cured, the NOA layer is removed from  the PDMS mold. Both components are then placed in a plasma cleaner for one minute. This primes their surfaces so that they are able  to adhere to each other. Finally, the body and base of the chip are pressed together. The base sheet can deform slightly during this step, especially around areas free of microfluidic structures, such as pillars. However, having such a flexible component in the chip is necessary to guarantee uniform bonding throughout the sample. The bonded chip is then exposed to UV light for 10 minutes, in order to complete curing. Freshly cured NOA is yellow, but its color is lost after exposure to white light. Thus, after fabrication we expose the chip to white light for about 24 hours, in order to stabilize its color for image analysis. This also stabilizes its wettability~\cite{DavideNOA}.

\subsection{Sample design}
\begin{figure}
\includegraphics[width=100 mm]{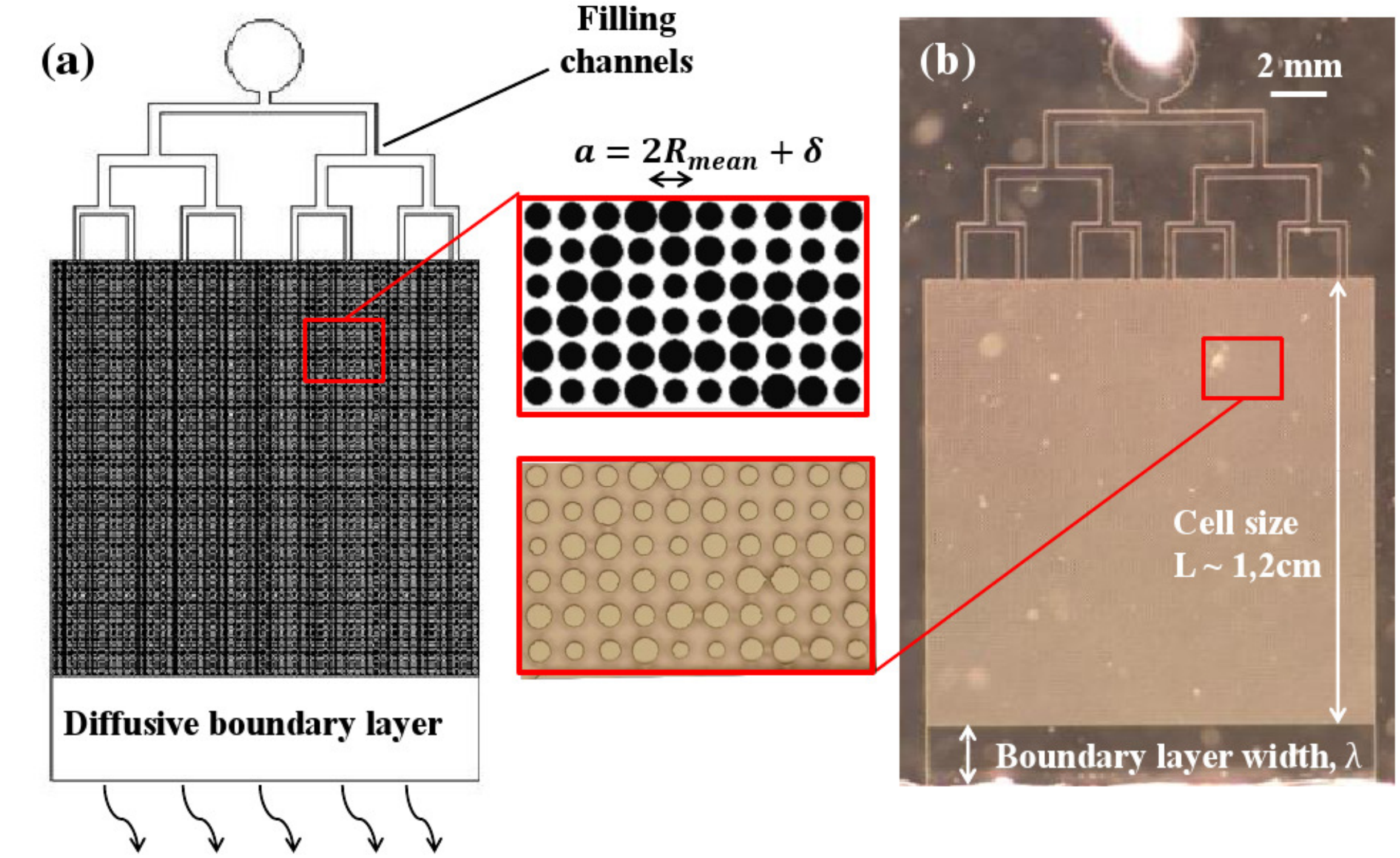}
\caption{From design to fabricated sample. The mask design (a) shows how a chip is conceived, with an inlet splitting into several channels that allow for a uniform filling of a square pore space. The actual sample (b) reproduces the design and the boundary layer, across which evaporation occurs, is cut to the desired width. The magnified regions show a direct comparison of the designed and realized features in the pore space.}
\label{sample}
\end{figure}

 After fabrication our samples are thin square cells with an open boundary layer on one side to allow evaporation, as can be seen in Fig.~\ref{sample}. We usually fix this layer to be $\lambda$ = 2 mm wide. However, to test the effects of different boundary layers on drying (e.g. \cite{Shahraeeni}), it could be varied between 0.5 and 4 mm. On the opposite side of the chip we place an inlet that successively splits into eight channels. We use these to fill the cells uniformly with a perfectly wetting, volatile oil or water. A 2D porous material is then realized by having an extensive array of round pillars in the cell. We can vary the sizes and positions of the pillars to mimic heterogeneous random packings, as would occur in a real soil. Using the soft lithography techniques described above, a minimum feature size of 5 $\mu$m and a feature resolution of 2 $\mu$m can be achieved. To simplify comparison to numerical modeling we designed our samples as an array of pillars lying on a square grid. Other designs, such as a random close packing or a triangular lattice, could just as easily be manufactured, to match models like those in Refs.~\cite{CieplakandRobbins,holtzman2015wettability}.

We introduce heterogeneity into the design by choosing the radii of individual pillars from some probability distribution. For example, the radius of each pillar could be randomly selected within the range of 45-55 $\mu$m (i.e. a uniform distribution with a relative width of $\pm$10\%).  We present experiments where the pillar radii heterogeneity within any one sample are randomly taken from a uniform distribution with a relative width of $\sigma_r$ = 3, 5, 10 or 20\%. Samples have a thickness of 40 $\mu$m, and contain a 100$\times$100 grid of pillars with an average pillar radius of $R = 50$ $\mu$m and average throat size of $\delta$ = 25 $\mu$m (and hence a grid spacing $a$ = 125 $\mu$m). Using these parameters we also avoid any overlap of pillars and maintain a minimum throat size of at least 5 $\mu$m. These designs give square cells of size $L \simeq1.2$ cm. 

Samples with correlated disorder have been produced as well. These will be described and presented in detail in a companion paper \cite{Borgman2016}. Here, we focus on the samples with uncorrelated disorder.
 
\subsection{Setup and image analysis}
To observe drying we place a cell filled with Novec 7500 engineered fluid -- a fluorinated oil supplied by 3M -- under a digital SLR camera (Nikon D5100) equipped with a macro lens. This allows for time-lapse imaging with a pixel resolution of 5 $\mu$m and a temporal resolution of up to 1 s. The sample is leveled and horizontal, in order to avoid gravitational effects on the drying process. The chip is also surrounded laterally by a ring of LEDs. The low-angle lighting allows the camera to see essentially only the light scattered from interfaces, in a technique similar to dark-field microscopy. The wet area of the chip therefore appears darker than the dry area, as the refractive index of the oil is intermediate between those of air and NOA. By darkening all other lights in the room we further ensure that there are no reflections from the top of the cell that might otherwise interfere with image analysis.

\begin{figure}
\includegraphics[width=145 mm]{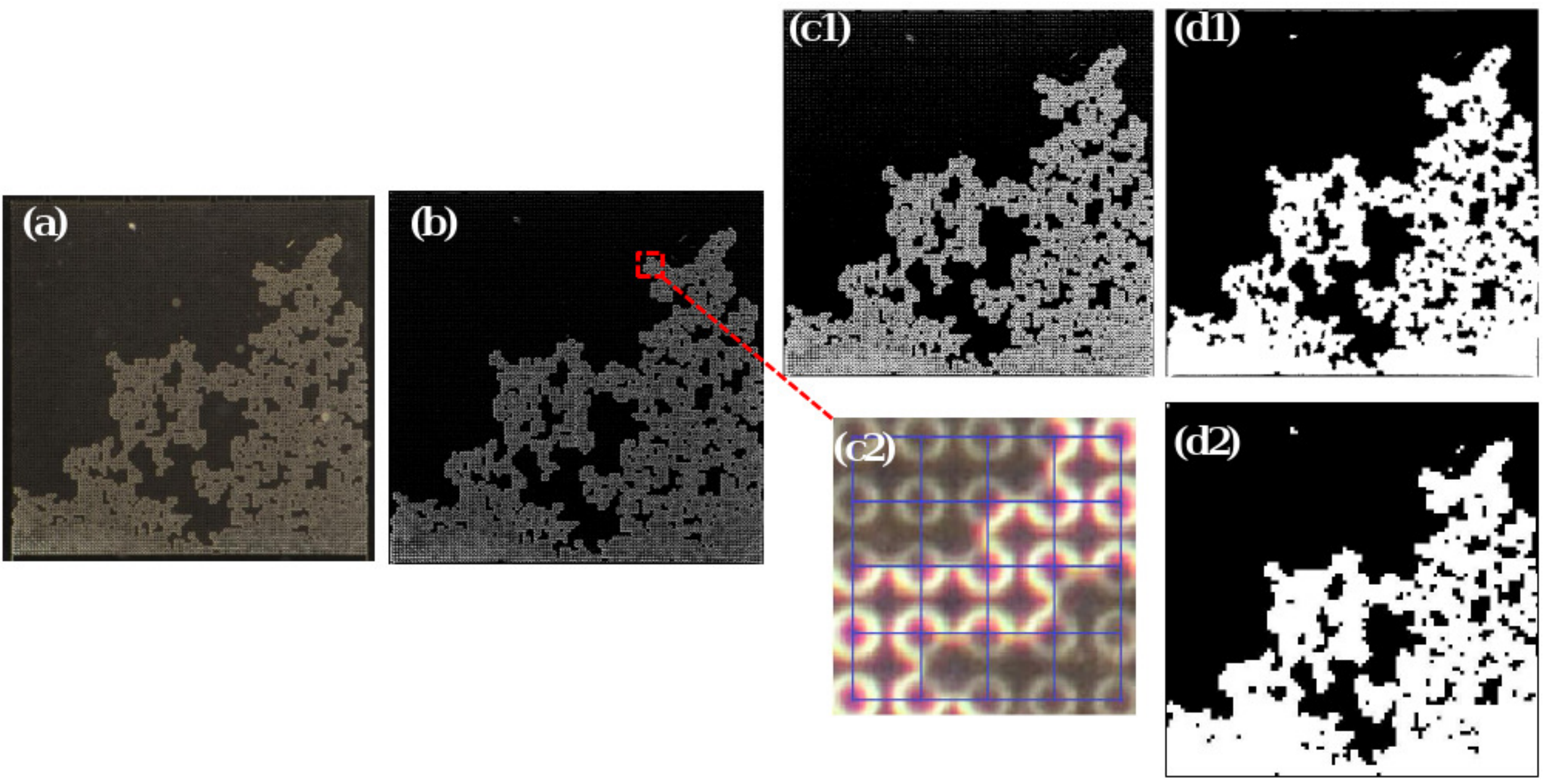}
\caption{Image analysis. We start from a color picture (a) which we flatten into a greyscale image and then apply a band-pass filter to eliminate both noise and intensity variations. Further subtracting constant sources of noise gives us a cleaned image (b). This can be either thresholded (c1) and transformed using morphological operations to produce a continuous black-and-white picture of the drying pattern (d1) or it can be discretized using its mask design and the intensity in each pore (c2) in order to show which specific pores are open at the given time (d2).}
\label{imanalysis}
\end{figure}

Once in place, the sample is allowed to dry while the camera takes a picture of it every minute. The time-lapse image sequence continues until the leading front of the oil-air interface reaches breakthrough, that is, when the drying front first reaches the filling channels, as this disrupts the invasion pattern. The resulting sequence of images is then cleaned and processed with Matlab. An example of this is shown in Fig.~\ref{imanalysis}. We start by extracting the red color channel of each image, which contains the best contrast. We then apply a bandpass filter to remove both the high-frequency noise (cutoff: 3 pixels) and any low frequency variations in intensity (cutoff: 30 pixels). From each image we also subtract the first picture in its sequence, in order to remove constant sources of background noise, such as dust or flaws in the microfabricated chips. Each cleaned picture (Fig.~\ref{imanalysis}b is then thresholded to give a binary image (Fig. \ref{imanalysis}c), that distinguishes between wet (black) and dry (white) areas. Next, we remove any white objects smaller than a pore size (about 100 pixels), such as wet pillars, which might still be partially visible in the binary image, in order to leave the wet area completely black. We then dilate the remaining white objects, affecting the contour of  the dry pillars and the air-liquid interface, and remove the isolated black regions corresponding to pillars in the dry areas.  Finally, we erode the picture to reverse the dilation and to give a map of the wet and dry regions where all the pillars have been removed. The accuracy of these image processing steps was monitored, and they were adjusted slightly, to prevent the loss of fine detail in any particular image sequence.

In order to exploit the precise knowledge of the geometry of our samples and to make comparisons with the PNM, we also employed a second stream of image processing, which summarizes the entire image sequence in a discrete matrix form. This is a 99$\times$99 matrix, $T_{ij}$, where each element specifies the time at which air first invades the corresponding pore, at location $(i,j)$.

To extract this matrix from our data, we start from the cleaned greyscale set of pictures mentioned above (Fig.~\ref{imanalysis}b). From this we manually find the pixel coordinates of the pores at opposite corners of the cell, corresponding to locations (1,1), and (99,99). These are then used to scale the mask design to the size of our images and thus to map all pillar coordinates. A pore is then defined as the open space between the centers of its adjacent four pillars. When looking at the area of the image that is around a single pore, there are four quarter-pillar crowns visible at its corners. We consider a pore dry when these crowns are dry. For any pore all four quarter-pillars were either dry or wet at the same time, as a partially filled pore was neither observed experimentally nor is allowed for in the model. A dry pore shows brighter crowns at the corners than a wet pore, as an air-NOA interface scatters light more than an oil-NOA interface does. We exploit this difference to choose an intensity threshold below which the pore will be considered wet. By analyzing the time-lapse sequence, we are thus able to determine the first time at which any particular pore ($i,j$) is observed to be dry. The pore invasion matrix, $T_{ij}$, records these times, and is demonstrated in Fig.~\ref{imanalysis}d.2.

\subsection{Scaling}
In order to simplify the description of our experiments and to ease comparison with the numerical pore-network model, we non-dimensionalize all variables. To do so, we first notice that the drying rate is constant in the early stages of our experiments. We find this drying rate by computing a numerical derivative: 
\begin{equation}
\dot{e}= \frac{1}{L}\frac{\Delta A}{\Delta t}
\end{equation}
where $\Delta A$ is the increase in dry area from one picture in the sequence to the next, $\Delta t$ is the  time interval between the two pictures and $L$ is the length of the side of the cell. The initial drying rate, $\dot{e_0}$, is taken to be the slope of a linear fit of the evaporation rate data over the first 40 minutes of any experiment.

We now define a characteristic length-scale using the spacing between adjacent of pillars, $a$. For example, the side of our cells has length $L=100a$, and so a dimensionless size of $L^*=100$. The initial drying rate is then used to define a characteristic timescale, $\tau$, as the time it takes to dry the first row of pores:
\begin{equation}
{\tau}=\frac{a}{\lvert\dot{e_0}\rvert} \rightarrow t^{*}=\frac{t}{\tau}
\end{equation}
where $\dot{e_{0}}$ is the initial drying rate and $a$ is the grid spacing of the pillars. If the drying rate was constant throughout the experiment, then $t^{*}=100$ would be the dimensionless time needed to dry the whole cell. 

\subsection{Minkowski functionals}
We want to quantitatively compare experimental and simulated patterns. In order to make rigorous comparisons, we use the Minkowski functionals to describe our invasion patterns \cite{Minkowski}. These metrics can be used to characterize all kinds of complex patterns arising, for example, from dewetting phenomena or fracture \cite{MeckeMinkowski, RalfKarinMinkowski}. In two-dimensional systems, three functionals are needed to characterize a pattern. These are: (\textit{i}) the ratio of one phase to the total area available, e.g. the liquid saturation, $S$; (\textit{ii}) the ratio, $\alpha$ of this area to its perimeter and (\textit{iii}) the Euler number, $E$. The last metric is related to the topology of the pattern, and gives the difference between the number of connected regions and the number of holes within them. These three metrics can thus give us information about the filling state of the system at a given time ($S$) the roughness of the drying front ($\alpha$) and the connectivity of the dry phase ($E$).

\begin{figure}
	\includegraphics[width=125 mm]{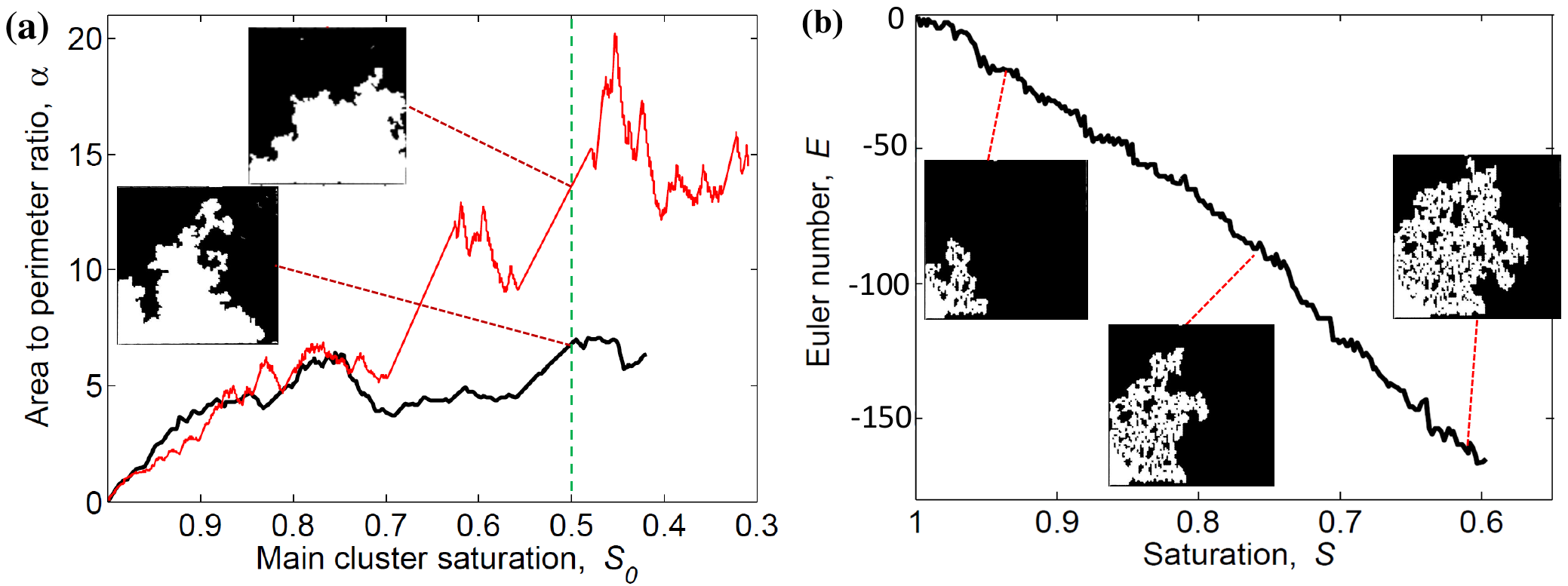} 
	\caption{Example of how the Minkowski functionals work. (a) The area to perimeter ratio, $\alpha$, plotted against the main cluster saturation, $S_{0}$, describes the evolution of the drying front in time. For a given invaded area, a smaller $\alpha$ means a higher perimeter and a rougher front, while an increase in the invaded area will show up as an increase in $\alpha$. Whenever a cluster is cut off from the main cluster, then both $\alpha$ and $S_0$ will be discontinuous and jump to new values. (b) The Euler number, $E$, is the number of connected objects (white areas) minus the number of liquid clusters (black areas). Therefore, a low $E$ means that there are many isolated clusters.}
	\label{AP example}
	\end{figure}

In our processed images the saturation, $S$, is the fraction of the total area covered by the liquid phase (the black pixels) divided by the total area of the porous medium. We note that the boundary layer and the filling channels are excluded from this calculation. In the case of $\alpha$, we consider only the shape of the leading front. This is defined by the air-liquid interface of the main cluster, which consists of all the wet pores that are connected with the reservoir at the back of the cell. The leading front has a perimeter length $P_0$ and thus divides the cell between the main, fully wet cluster, and a region of area $A_0$, which is either dry or contains isolated clusters of fluid. The fraction of the porous medium covered by the main cluster also defines the main cluster saturation, $S_0$. The ratio of the leading area and perimeter, $\alpha$, can be scaled to be non-dimensional using the characteristic grid size, $a$, as:
\begin{equation}
\alpha=\frac{A_0}{a P_0}.
\end{equation}
A high $\alpha$ therefore means a small $P_0$ and corresponds to a compact front. A rough front would have a lower ratio, as we show in Fig.~\ref{AP example}.

Finally, the Euler number is computed by counting the number of connected regions (the dry area, in our case) and subtracting the number of holes (all the isolated clusters). Therefore, the more negative the Euler number is, the more isolated clusters we have.

\section{Model}

Simulations of drying cells are performed using a pore-network model inspired by \cite{Ran2010,Prat1993}. The goal is to capture the main features arising in experiments using as simple a description of the pore-scale physics as possible, and a minimal set of rules for how pores interact. 

\begin{figure}
\includegraphics[width=75 mm]{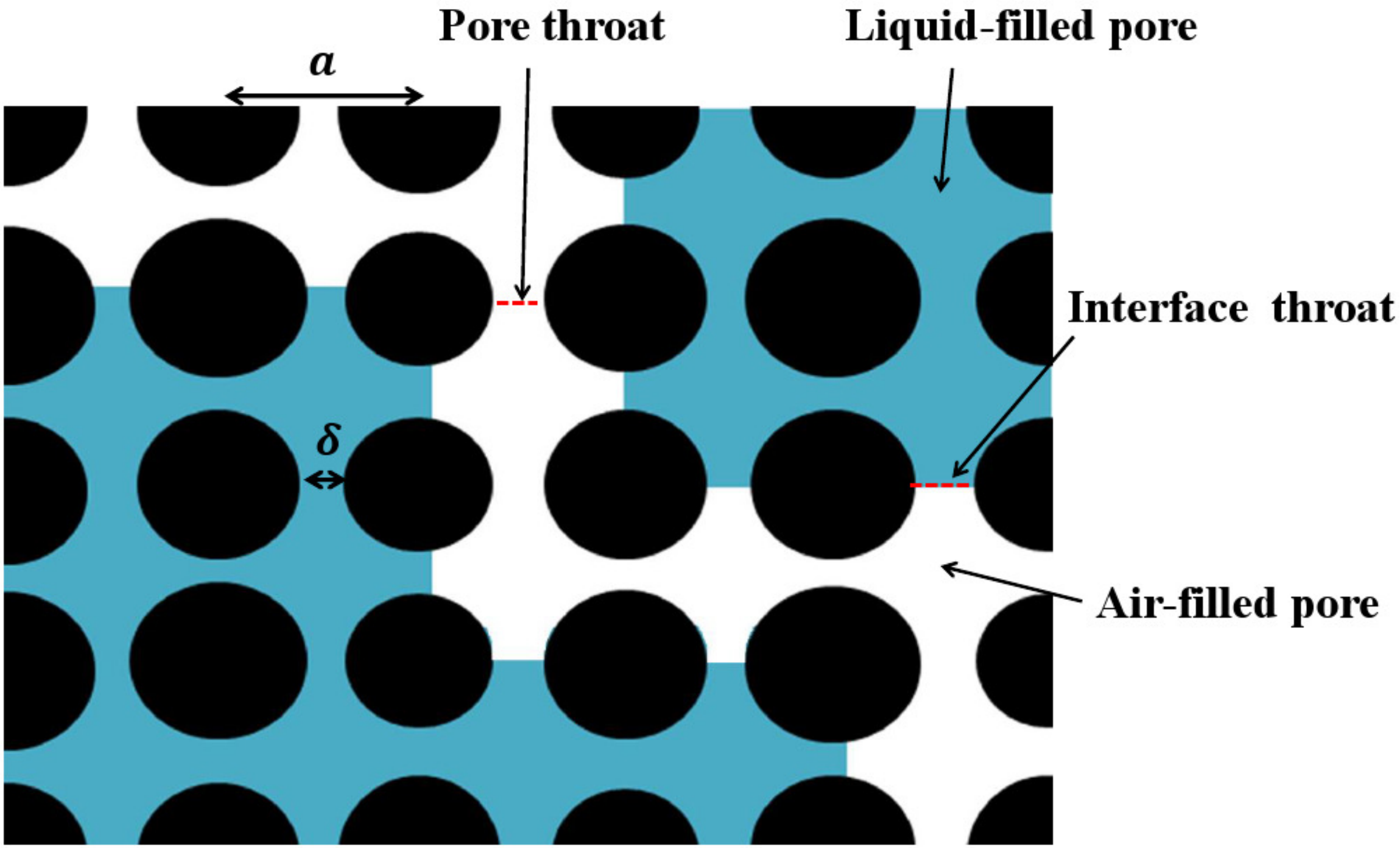}
\caption{Schematic description of the pore-network model, based on the geometry of microfluidic cells. Pillars (in black) of different radii are placed on a grid, with a lattice spacing $a$ between adjacent pillar centers. Pores may be either liquid filled (here, blue) or gas filled (white). Any two adjacent pillars, $i$ and $j$, are connected by a throat, of size $\delta = a - (r_i + r_j)$. Each connected cluster of pores has a capillary pressure $p$, that is related to the curvature of menisci in the throats along its fluid-air interface.}
\label{fig:model.schem}
\end{figure}

The experimental geometry is modeled as a set of discrete pores, which communicate through throats, or constrictions, as sketched in Fig.~\ref{fig:model.schem}. The sizes of both types of objects, pores and throats, are specified by the experimental design. Defining a pore as the space enclosed by four pillars, the pore volume is $V=(a^2-\sum_{i=1}^{4} {\pi r_i^2}/{4})h$ where $a$ is the grid spacing, $r_i$ is the radius of one of the surrounding pillars and $h$ is the thickness of the cell. Throats are the minimum distance between two adjacent pillars. For example, if these have radii $r_i$ and $r_j$, then the throat size is $\delta=a-(r_i+r_j)$. 

We assume that two timescales play a role in the drying process, namely those of vapor diffusion and pore invasion. As diffusion is much slower than pore invasion, we are able to separate these two timescales. In other words, we treat invasion events as effectively instantaneous. Between any two sequential invasion events we model the equilibrium vapor concentration, $\rho$, in the dried pore space, and use this to solve for the evaporation rate of the volatile phase -- here the evaporating oil. Boundary conditions for the diffusion problem are provided by assuming a saturated vapor density at the air-liquid interface, $\rho = \rho_{s}$, and that the vapor density at the edge of the diffusive boundary layer (i.e. the open atmosphere) is zero, such that $\rho = 0$ there.

Liquid is removed from the sample through evaporation, at a rate approximated by Fick's first law.  Between any two adjacent pores $i$ and $j$ there can be a vapor flux: 
\begin{equation} \label{eq:fickslaw}
q = - \ell D  \nabla \rho = \ell D \frac{\rho_i - \rho_j}{a}.
\end{equation}
Here, $q$ is the flux per unit cross-sectional area (i.e. the total mass transfer rate between the two pores is $q\delta h$), $\delta$ is the width of the throat between the two pores and $D = 5\times 10^{-6}$ m$^2$/s is the diffusion constant of the liquid vapor in air.  Furthermore, the difference in vapor concentrations at the centers of the two pores is $\Delta \rho$, and their center-to-center distance is $a$. The prefactor $\ell$ allows us to distinguish between diffusion in the porous medium, and in the boundary layer. Between any two pores in the chip diffusion is restricted by the throat width, and we set $\ell=\delta/a$. The boundary layer, of width $\lambda$, is instead modeled by an array of effective pores with $\ell=1$. We compute the vapor concentrations and fluxes in the medium by simultaneously solving for the mass balances on all pores. In a steady state this is equivalent to solving the continuity equation $\nabla\cdot q = 0$, or the Laplace equation for vapor concentration, $\nabla^2 \rho = 0$, in the continuum approximation.

Initially, our porous medium is fully saturated with fluid. The potential drying rate in this situation can be solved exactly to be:
\begin{equation} \label{eq:rate}
\dot{e}_{pot} = \frac{ D  P_{sat} m}{\lambda R T \rho_{liq}},
\end{equation}
where $P_{sat}=2.1\times10^{3}$ Pa is the saturation pressure of our oil, $\rho_{liq}=1614$ kg/m$^3$ is its liquid density, $m=0.414$ g/mol is its molar mass, $R$ is the universal gas constant and $T=298$ K is the lab temperature. The initial drying of the PNM reduces to this situation in the limit of small pores and a thick boundary layer ($\lambda /a \gg 1$).

\begin{figure}
	\includegraphics[width=100 mm]{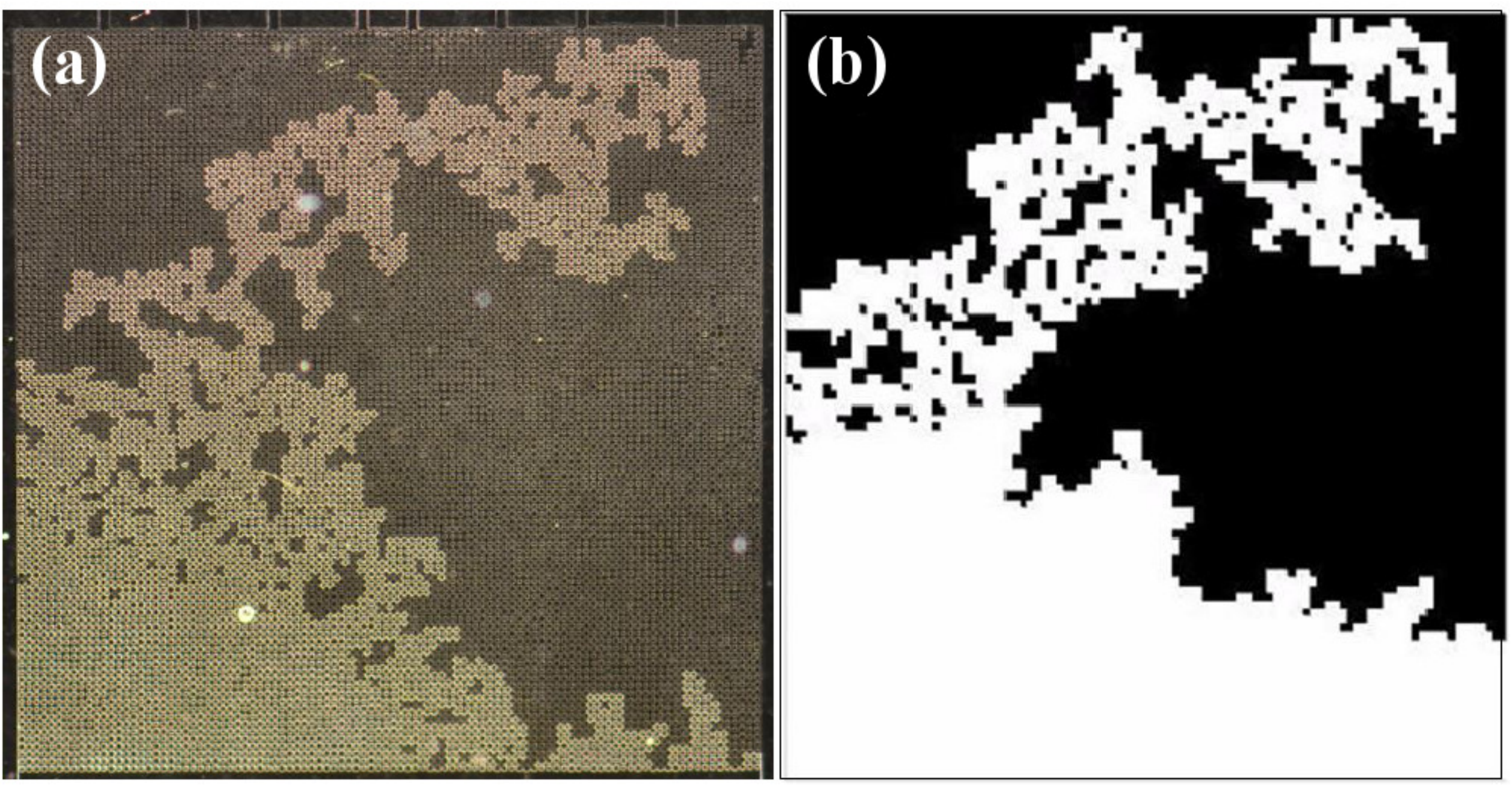}
	\caption{Comparison between experimental (a) and simulated (b) patterns at breakthrough using identical pore geometries.}
	\label{exp vs model}
	\end{figure}

In order to model pore invasion we approximate the air-liquid menisci of interfacial throats as ellipsoidal caps. These caps are characterized by a dynamic horizontal radius of curvature, limited by the throat size $\delta$, and a fixed, vertical radius of curvature, set by the cell's thickness, $h$. The pressure-volume relation in such caps is given in \cite{Shahraeeni}, and for this calculation we assume a perfectly wetting fluid phase. As evaporation occurs liquid is removed from the fluid-filled pores, causing their capillary pressure to increase, along with the curvature of the menisci in the interfacial throats. For every throat there is a critical pressure that is set by the Young-Laplace equation:
\begin{equation}
p_{c} = {\gamma}\bigg(\frac{2}{\delta}+\frac{2}{h}\bigg),
\end{equation}
where $\gamma=16.2$ mN/m is the air-liquid surface tension. Once the critical pressure is achieved at any pore along the fluid-air interface, then that pore is invaded and its liquid volume is redistributed amongst all the saturated pores connected with it. Although this is a simple model, we show in Fig.~\ref{exp vs model} that it can nonetheless faithfully capture many macroscopic details of the experimental drying patterns.

\section{Results}
This paper aims to present an experimental technique that allows unparalleled control over a pore-scale process and to bring this together with a simple pore-network model to test how well can it reproduce experimental data. In the next section we start by comparing absolute drying rates. We then proceed to test how reproducible our microfluidic experiments actually are and the corresponding sensitivity of our model. Finally, we compare the finer details of the patterns observed in experiments to those observed in the corresponding models, and assess the successes and limitations of the pore-network approach, with a view for future applications.  
	
\subsection{Drying rates}
An experiment begins with the onset of air invasion into the drying porous medium. In all of our  experiments we observe an initial phase (of at least 40 minutes) where the drying rate, $\dot e$, is constant within experimental uncertainty. The duration of this constant rate period depends mainly on the size of the boundary layer, as for early drying the pore geometry of the cell is expected to have little to no effect on the drying dynamics~\cite{Shahraeeni}.

\begin{figure}
	\includegraphics[width=125 mm]{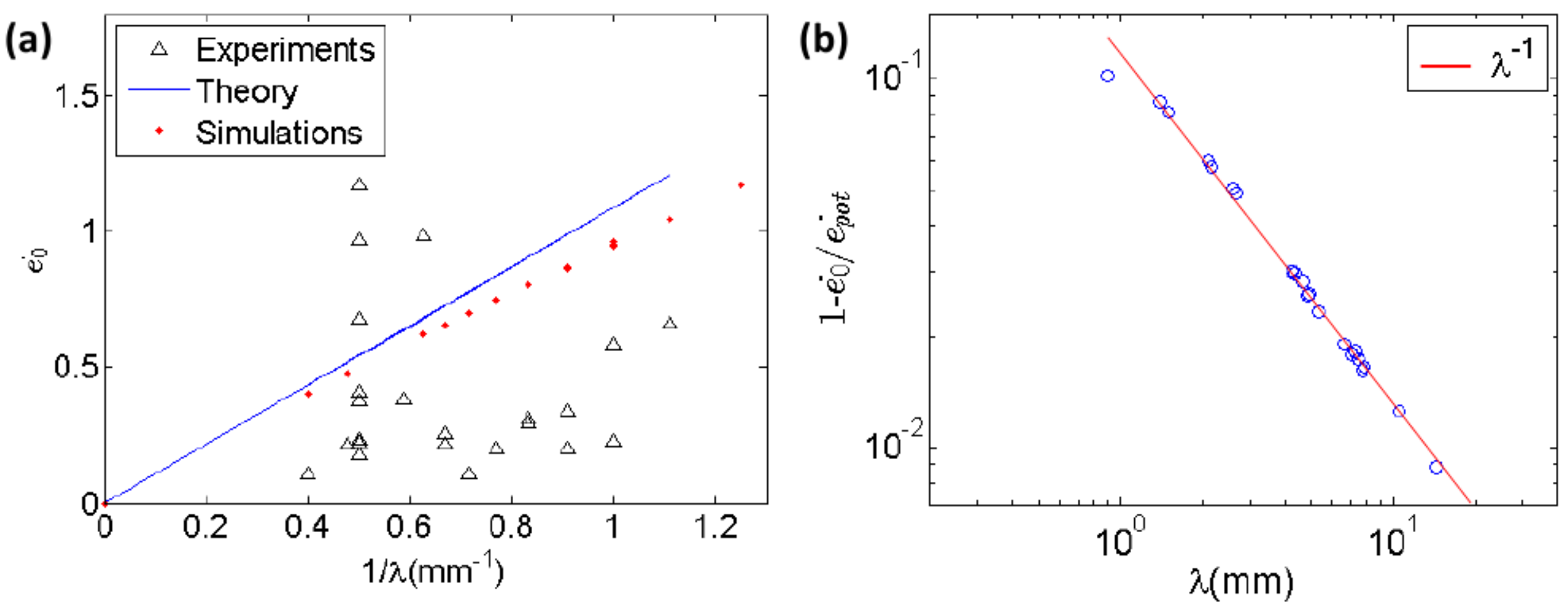}
	\caption{The initial, absolute drying rates should scale with the inverse of the boundary layer length, $\lambda$.  In (a) the blue dots refer to the simple theoretical prediction of Fick's law (Eq.~\ref{eq:rate}). The red dots show the simulated rates and the black triangles the experimental initial drying rates.  In (b) we show how the theoretical and simulated rates approach each other asymptotically, as reported in \cite{Shahraeeni}.}
	\label{e0 vs BL}
\end{figure}

The experimental evaporation rates have two main sources of error: variations in the height of the boundary layers that may be introduced while sealing the samples (e.g. bending of the substrate) and noise due to the discretization and measurement of the time-lapse images. Otherwise, experiments were carried out in well-controlled conditions, with the temperature being fixed to within $\pm 1$\textdegree C. Considering the measurement error first, we found that during the initial drying period the measured drying rate would naturally fluctuate with an average standard deviation of 0.12 $\mu$m $s^{-1}$ around the mean rate, in any particular experiment. We take this as our absolute measurement uncertainty. However, as shown in Fig.~\ref{e0 vs BL}, it is clear that the systematic variation in conditions between experiments is more important. In fact, the data at $\lambda^{-1}$~ $=$~ 0.5 ~mm$^{-1}$ show how the initial drying rates can vary greatly between experiments, even while keeping the boundary layer constant.

In Fig.~\ref{e0 vs BL} we show how the experimental and simulated initial drying rates, $\dot e_0$, depend on the inverse of the boundary layer's length, $\lambda^{-1}$. As well, we show the potential drying rate given by Eq.~\ref{eq:rate}. In the simulations the initial drying rate is extracted by solving for the vapor density in the boundary layer and then computing the vapor mass flux in the pores along the air-liquid interface. For evaporation off of a pool of fluid the drying rate should scale inversely with the boundary layer's length, as described in Eq.~\ref{eq:rate}. For evaporation over a wet porous body, $\dot e_0$ is expected to be slightly lower, by an amount that depends on the ratio of the pore size to the boundary layer width and the relative coverage of pores at the surface \cite{Shahraeeni}. However, the ratio of drying rates from a free fluid and a porous surface should asymptotically approach one as the boundary layer increases: such behavior is demonstrated in Fig.~\ref{e0 vs BL}b.

When compared to experiments, the simulations capture the right order of magnitude of $e_0$, although experiments typically show a lower initial drying rate (about half of that expected). This discrepancy between model and experiment can be reduced if we consider an additional effective boundary layer of about $3-8$ mm. The difference could thus be accounted for by the existence of a small additional effective boundary layer in the stagnant air outside of the microfluidic chip, with additional (and unmodeled) gradients in the vapor concentration there.

As we will show in the next sections, however, uncertainties in the effective boundary layer width affect the initial drying rates alone and will not significantly change the patterns formed during drying. Therefore, in order to simplify further analysis and comparisons, we will use these initial drying rates to scale our observations as described in the \textit{scaling} section. 

\subsection{Repeatability}

Next, we examine the reproducibility of the drying patterns, experimentally. A source of uncertainty in the experiments is the manufacturing of the cells. Therefore, we want to know: how well does the pillar size distribution in a chip match its design? Then, we estimate how much this error influences a drying pattern, in order to answer the question: will the metrics stay the same for different copies of the same master design?
To answer the first question, we compared digital microscope pictures of our samples with the designs of their masks, as sketched in Fig.~\ref{sample}. Measuring the size of 200 pillars gave an average radius of $49.72\pm 0.18~\mu$m, compared to a designed size of 50 $\mu$m. Thus, we have a negligible systematic error in manufacturing, of at most $0.28 \pm 0.18$ $\mu$m. However, when measuring 50 single pillars and comparing them to their design, we have found their radii to be, on average, $1.63 \pm 0.20~\mu$m different from the original design. This corresponds to a 3.2\% random manufacturing error in feature size. During manufacture we also measured variations in the thickness of the pore space by means of a white light interferometer. Within each sample, as measured at the four corners, we tolerated variations in thickness of no more than 3 $\mu$m. For this paper all samples were, on average, 38 $\mu$m thick. 

\begin{figure}
	\includegraphics[width=120 mm]{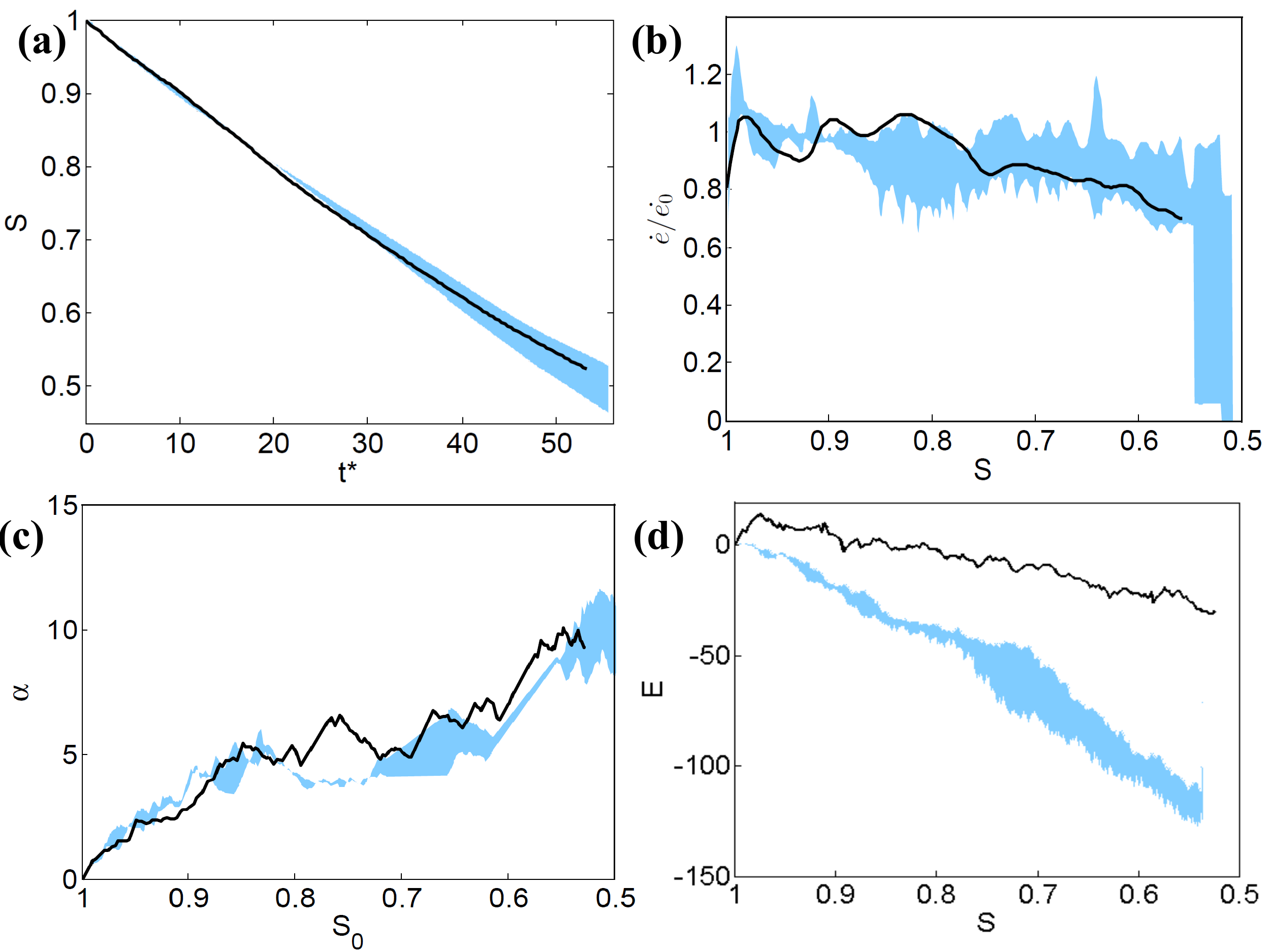}
	\caption{Metrics showing the experimental clouds of observations, based on three replicates of identically prepared chips (the blue area gives the mean behavior within one standard deviation). The black line corresponds to an experiment in the same design, but with water, instead of oil, as the volatile phase. Shown are:  Saturation vs. (scaled) time (a);  Drying rate vs. saturation (b);  the area-to-perimeter ratio vs. main-cluster saturation (c);  and the Euler number vs. saturation (d).}
	\label{repeat+water}
	\end{figure}
	
The next question was: how reproducible are the drying patterns? We have repeated experiments on three designs in order to establish the expected size of the experimental cloud of results. We did this by using, for each design, three different chips cast from the same master. We measured the Minkowski metrics during drying in each case and estimated a reproducibility cloud for our experiments. The boundaries of this cloud are established by averaging the results of the three samples and calculating their standard deviation. An example of the experimental cloud thus established can be seen in Fig. \ref{repeat+water}, where the blue area shows the range of our tests. We can see how the same design tends to follow the same path, despite minor experimental variations. The black line in Fig.~\ref{repeat+water} shows the results of an additional experiment performed using water as the volatile phase, and which will be discussed in a later section.

To determine how well the exact drying patterns can be reproduced we also compared experiments pore-by-pore using the matrices shown in Fig.~\ref{imanalysis}d.2. In order to do this we first used the pore invasion matrix, $T_{ij}$, to find which pores had been invaded at any time $t_0$ as
\newcommand*{\bfrac}[2]{\genfrac{}{}{0pt}{}{#1}{#2}}
\begin{equation}
A_{ij}= \Bigg\{\ \bfrac{1, \ \ \  ~T_{ij}\leq t_0}{0, \ \ \  ~T_{ij}> t_0}.
\end{equation}
We then confined our attention to the invasion pattern of the main cluster by removing from $A_{ij}$ all isolated clusters, or any isolated patches of ones. At the same main cluster saturation, $S_0$, we could then compare two invasion patterns $A$ and $A'$ by computing their overlap, or match
\begin{equation}
\Delta= \frac{A_{ij}\cdot A_{ij}'}{N}
\end{equation}
where $N=\sum_{i,j}A_{ij}$. In other words, we find the fraction of invaded or isolated pores, $\Delta$, that match each other in both patterns, at the same main cluster saturation.

\begin{figure}
	\includegraphics[width=120 mm]{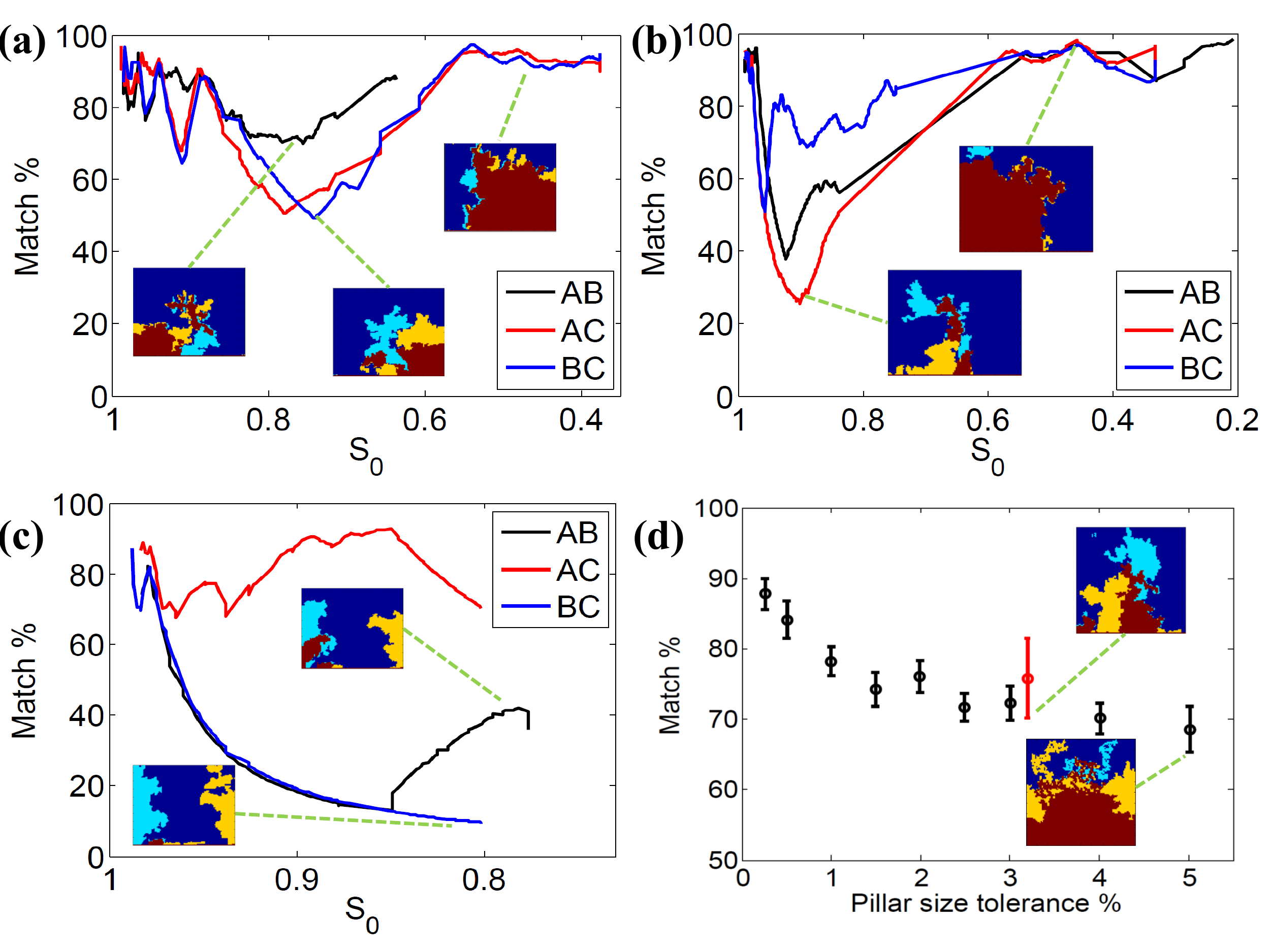}	
	
	\caption{Pore-by-pore comparison of the leading patterns during drying. We repeated three experiments for each of three separate designs. In panels (a) -- (c) we show how well the drying patterns match each other between pairs of replicates. Insets show the differences in invasion patterns. The colors highlight areas that are unsaturated in both samples (red), in one sample only (yellow or light blue) or part of the main cluster in both samples (dark blue). In (d) we also show how increasing the manufacturing error in our designs decreases the achievable match in the leading patterns of numerical simulations, when $S_0$ = 0.5. The red point compares this with the average agreement between our experiments and their corresponding simulations, again when $S_0$ = 0.5 (see Fig. \ref{pore-pore}).}
	\label{pore-pore exp}
	\end{figure}

The evolution of $\Delta$ is shown in Fig.~\ref{pore-pore exp} for each of our three replicated experiments. Figure \ref{pore-pore exp}a,b shows how the invasion patterns in replicates can be reproduced with a match of up to $\Delta \simeq$90\%, and typical values of about 80\%. In contrast, Fig.~\ref{pore-pore exp}c shows a case where the similarity of the patterns is lower than $\Delta=10$\% at breakthrough. However, the insets explain why: drying in this particular sample reached an early binary choice, based on the near-surface pore sizes, and evolved the drying front either on the left (replicates A, C) or the right (replicate B) side of the cell, hence the lower $\Delta$. 

In order to predict the agreement that could be expected between experiments and numerical modeling, we also tested how a random manufacturing error may affect the invasion patterns during drying. For this, we ran a series of numerical simulations with the same initial geometries (radius distributions with relative width $\sigma_r=0.2$), but with additional random perturbations in the pillar sizes. Figure~\ref{pore-pore exp}d shows the match that we were able to achieve between different numerical simulations when introducing such a random error. There, we see how increasing the manufacturing error gradually reduces the possible match between patterns, going down to $\sim$70\% when as little as a 5\% error was introduced.

\subsection{Sensitivity analysis}
\begin{figure}
\includegraphics[width=120 mm]{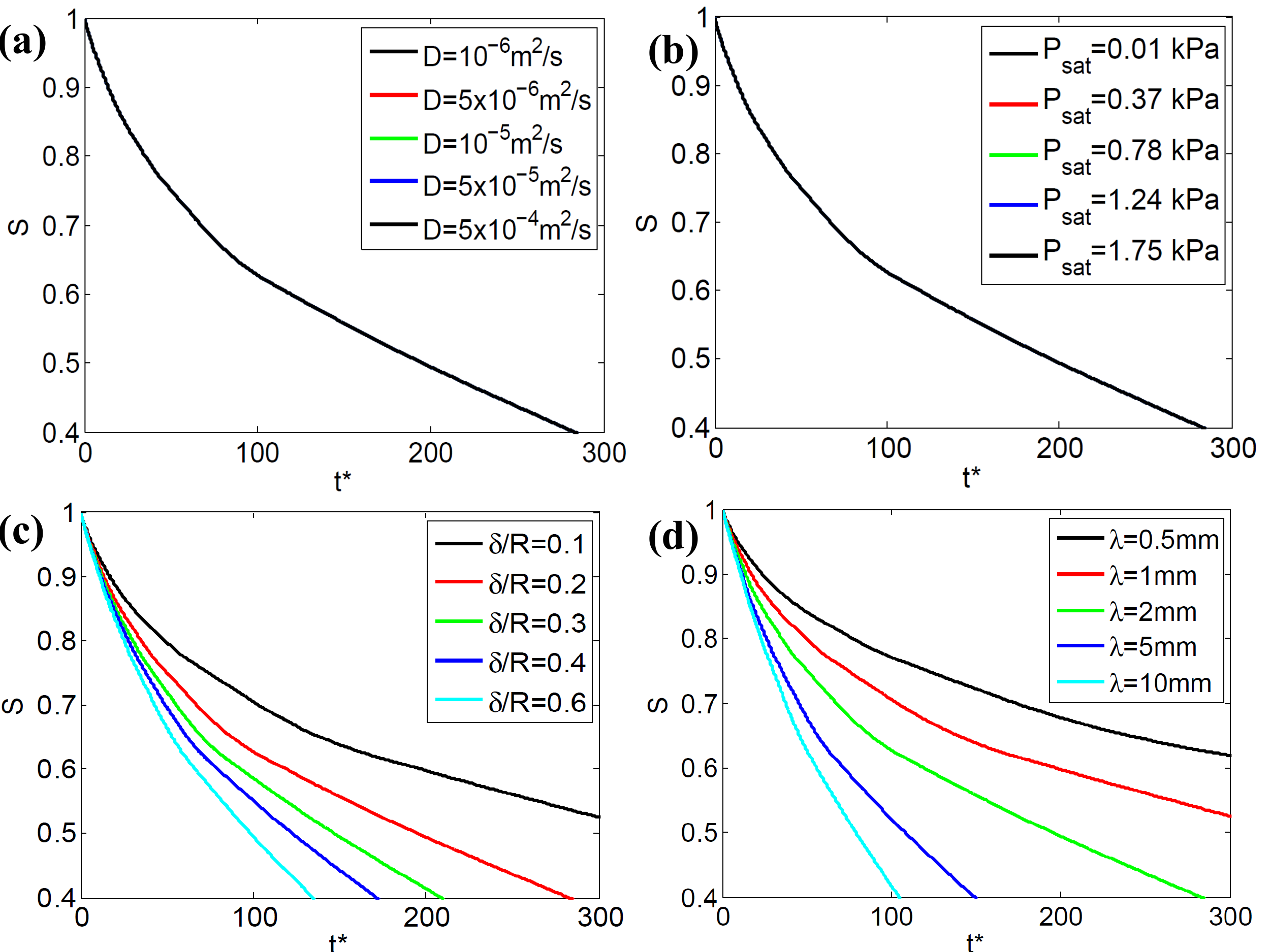}

\caption{We show how the saturation-time curves vary when using a time $t^*$ scaled by the initial evaporation rate, as parameters change in our model. The parameters are the diffusion coefficient, $D$ (a), the vapor saturation pressure $P_{sat}$ (b), the throat aperture-to-pillar radius ratio, $\delta /R_{mean}$ (c) and the boundary layer's width, $\lambda$ (d). When varying $D$ and $P_{sat}$ the scaled drying curves collapse. However, changing the mean throat size or the boundary layer's width has an effect on how quickly drying slows down over time.}
\label{fig:sensitivity}
\end{figure}

Here, we use simulations to evaluate the sensitivity of our system to several control parameters. Some parameters, like throat aperture and boundary layer length, are easier to control experimentally than others, like temperature, air pressure and relative humidity. In order to check that our scaling correctly accounts for uncertainties in the room conditions, while leaving in all geometrical effects, we chose one particular experimental geometry to model, then individually varied (i) the vapor diffusivity $D$, (ii) the vapor saturation pressure $P_{sat}$, (iii) the relative mean throat size $\delta/R$ and (iv) the boundary layer width $\lambda$ and observed how the saturation-time curve changed. The results of this sensitivity analysis are shown in Fig. \ref{fig:sensitivity}.

From Fig.~\ref{fig:sensitivity}a-b it is clear that $D$ and $P_{sat}$ have no additional effect on drying beyond the scaling of Eq.~\ref{eq:rate} and that this effect is accounted for entirely by our choice of dimensionless time. However, changing the throat aperture or the boundary layer's thickness does result in different behavior. 

Changing the throat aperture changes vapor transport properties in the model, as shown in Eq.~\ref{eq:fickslaw}. In Fig.~\ref{fig:sensitivity}c we show how decreasing $\delta/R$, and, consequently, offering more resistance to the vapor transport within the medium (as compared to the surface boundary layer), causes the drying rate to slow down more, as the drying front recedes into the porous medium. This effect is not surprising, as it proves how the throat sizes regulate transport within the porous medium.

Changing $\lambda$ also causes the drying rates to slow down at different paces. Specifically, a thicker diffusive boundary layer effectively separates the porous medium from the atmosphere by a long diffusion path. When this path remains large, as compared to the diffusive path through the pore space, evaporation will stay roughly constant. Conversely, for a short boundary layer, the drying rate will vary more as the drying front recedes from the surface.

\subsection{Disorder and drying behavior}

\begin{figure}
	\includegraphics[width=120 mm]{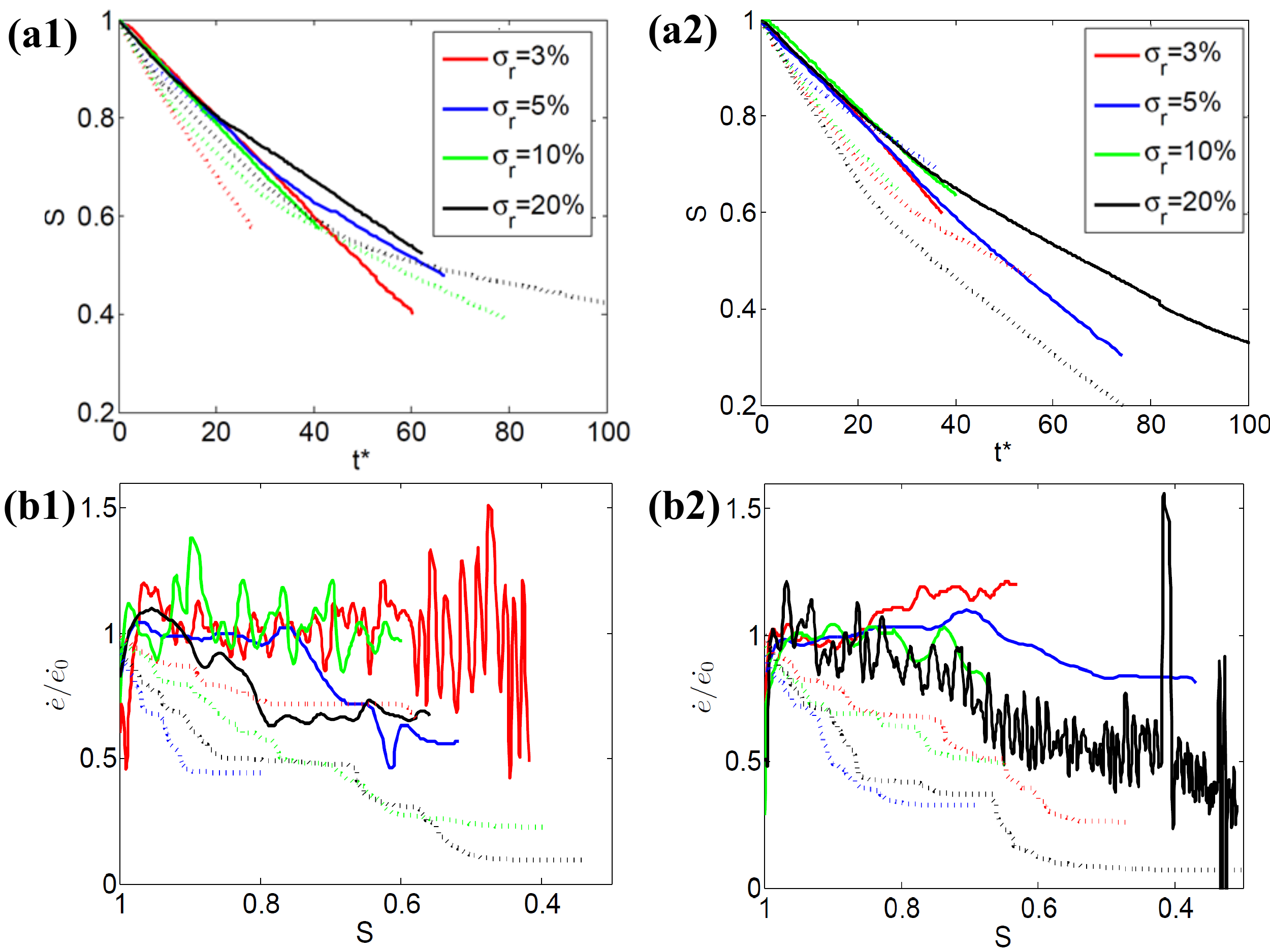}
	\caption{Saturation (a) and drying rates (b) of samples with varying degrees of disorder. The left column (1) shows one set of experiments, while the right column (2) shows a re-randomization of the same experiments. Solid lines refer to experimental results, whereas dotted lines (of the same color) show corresponding numerical simulations, using the same pillar sizes.}
	\label{sat_rate disorder}
\end{figure}

Having characterized the variations expected in both experiments and simulations, we will now compare the two cases to each other directly via their drying rates and Minkowski metrics. For this we also changed the amount of disorder in our cells, in order to test the effects of local heterogeneity on drying. As discussed in the methods section, we generated distributions of pillars where we allowed the radii to change within a 3, 5, 10 and 20 \% window. Two different randomizations of the pillar radii were made for each level of disorder, with different random seeds. The same exact geometries were then reproduced in our simulations. In Fig.~\ref{sat_rate disorder}a we first compare drying curves as a function of the dimensionless time $t^{*}$ and then the relative evaporation rates as a function of the liquid saturation, $S$. In each plot we compare experiments (solid lines) and simulations (dotted lines). Lines of the same color within the same plot refer to the same sample geometry.  

	\begin{figure}
	\includegraphics[width=120 mm]{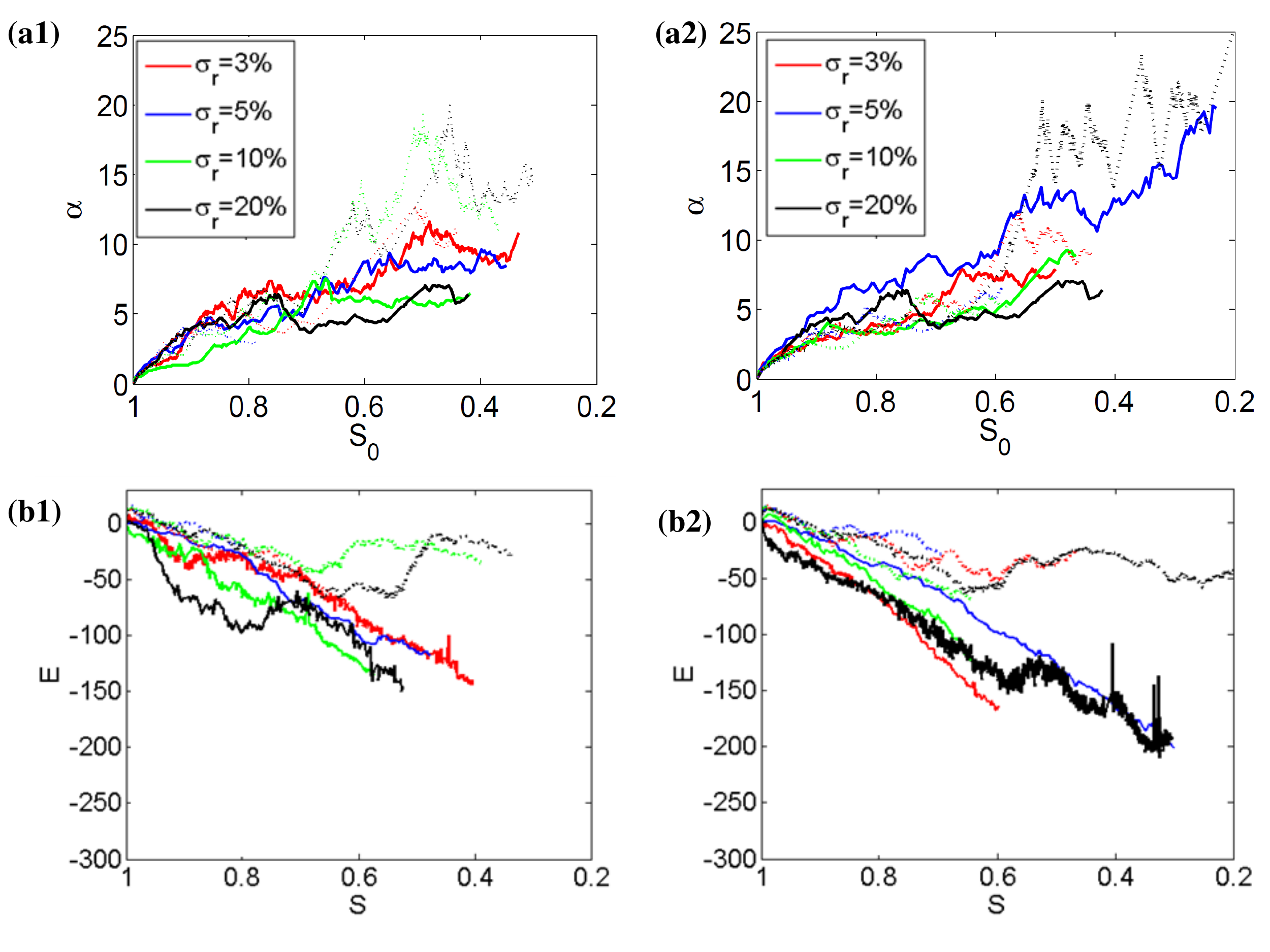}
	\caption{Minkowski functionals for the two different re-randomizations (columns (1) and (2)) and levels of disorder. Panels (a) show the area-to-perimeter ratio, $\alpha$, vs. main cluster saturation $S_0$ and (b) show the Euler number, $E$, vs. saturation, $S$.  Line colors and types are as in Fig. \ref{sat_rate disorder}.}
	\label{MM}
	\end{figure}

There was no clear effect of disorder on either the saturation-time curves or on how the evaporation rates of the samples evolve. However, we observed that experiments would always start by keeping a constant rate period and that, in some instances, the beginning of the falling rate period can be seen before breakthrough. In contrast, the simulated evaporation  rates always drop immediately. They usually then stabilize around a relative value of 0.4-0.8, and later drop again. In order to understand the origin of this discrepancy we further compare drying patterns by using the Minkowski functionals. 

We show how these functionals evolve during drying in Fig.~\ref{MM}. The roughness, $\alpha$, of the leading front, as a function of the main cluster saturation, $S_0$, is reproduced well in the simulations, increasing in a very similar way to that observed experimentally. However, the dynamics of the Euler number, $E$, are not reproduced as well. In the experiments, the receding drying front leaves behind large numbers of clusters that evaporate very slowly. In contrast, fewer isolated clusters form in the simulations and these clusters disappear faster than in the corresponding experiments. The difference in the behavior of isolated clusters could explain the discrepancy observed in the experimental and simulated drying rates. For example, the persistence of the isolated clusters in the experiments effectively maintains higher drying rates by increasing the vapor concentration and enhancing transport within near-surface pores. We will test this idea further in the next section.

\begin{figure}
	\includegraphics[width=135 mm]{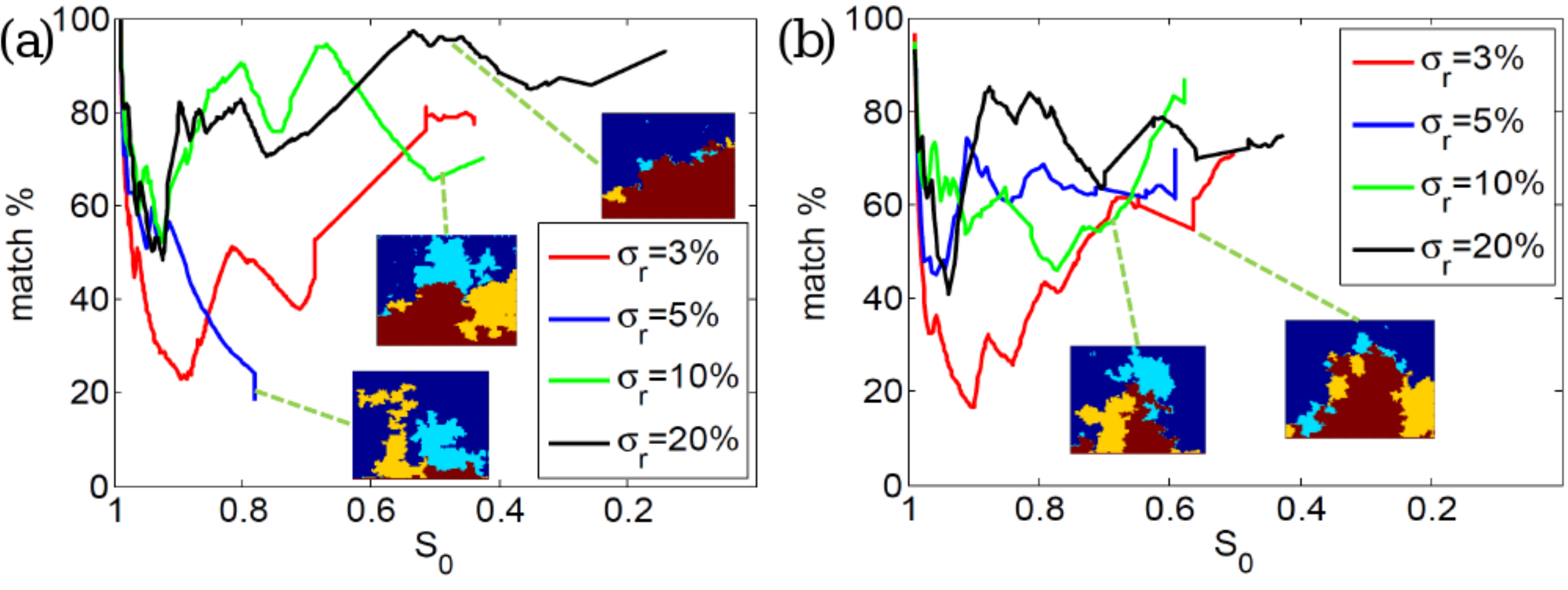}
   	
	\caption{Pore-by-pore comparison of leading patterns, between paired experiments and simulations. The plots show the percentage difference or match, $\Delta$, between the two cases, versus $S_{0}$. Shown are two different randomizations (a), (b).  The insets show the overlap between the experimental and simulated invasion patterns at particular values of $S_0$; their color code matches that in Fig. \ref{pore-pore exp}.}
	\label{pore-pore}
	\end{figure}

Since the patterns affect the drying rates, we have quantified their agrement more precisely by making a pore-by-pore comparison between experiments and simulations, as we did when comparing experiments in Fig. \ref{pore-pore exp}. This result is summarized in Fig.~\ref{pore-pore}. There we see how the agreement in most cases stays within, or at least close to, the limit of $\Delta$~=~70\%. This is a reasonable result when compared to the $\Delta$~=~90\% threshold at breakthrough we established in the \textit{Reproducibility} section, when comparing pairs of identically prepared experiments. Indeed, in figure \ref{pore-pore exp}d we actually show how a manufacturing error of 3\% (as we measure in our samples) brings the best expected agreement into the 70-80\% range. Therefore a 70\% match is as close to a perfect agreement as it is possible to achieve, given our experimental tolerance. There is one exception where $\Delta$ quickly decreases to a value lower than 30\%, and does not increase again before the front reaches breakthrough.  Although this case is not as marked as that in Fig.~\ref{pore-pore exp}c, we can see from the insets in both figure panels that here the front evolves as a consequence of an early binary choice over which part of the cell to invade. This is a demonstration of the high sensitivity of the invasion pattern on the exact details of the sample geometry. Making even a single pore smaller can, in fact, force the front to evolve very differently from the way predicted by the simulations.    
	
In summary, in neither our experiments nor our numerical model was there any significant effect of the magnitude of random disorder in grain size on the invasion patterns.  In most cases we showed that the leading patterns are captured excellently by the model, especially when the pattern agreement is considered in light of the achievable manufacturing precision.  However, the model and experiments disagreed on the Euler number (or the density of isolated clusters trapped behind the leading front) and the drying rates.  These quantities are, in fact, related and in the next sections we investigate possible reasons for the remaining discrepancies that we observe.

\subsection{Diffusion versus Invasion}

\begin{figure}
	\includegraphics[width=80 mm]{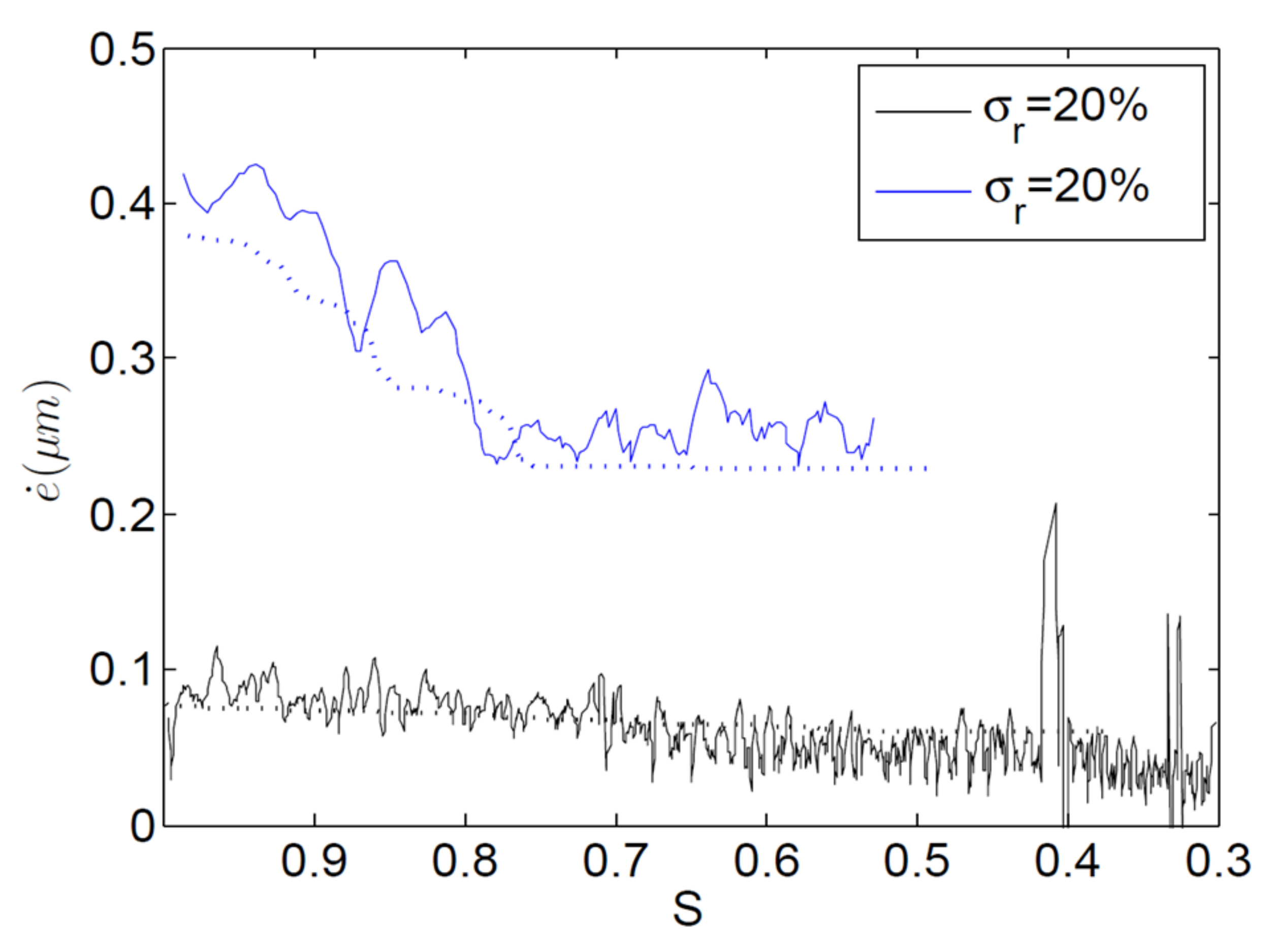}

	\caption{Absolute evaporation rates calculated by modeling Laplace's equation for the experimentally observed sequence of drying patterns, for two different trials. Experimental rates are shown as solid lines, while simulated rates are shown as dashed lines. }
	\label{from sep}
	\end{figure} 

The different behavior of drying rates in experiments and simulations can be explained by considering the different way in which isolated clusters form and behave. In the model, these isolated clusters of liquid evaporate very quickly, whereas they are less active in experiments, often persisting until breakthrough. If we take the experimentally evolved patterns, at various times, we can use our numerical model to predict what the resulting evaporation rate should be, for those exact patterns.  By doing this for each image in the sequence of a drying experiment, we can test whether the difference between the observed and predicted creation and loss of clusters can explain why the simulated evaporation rates behave differently from the experiments.  In the process, we can also implicitly test whether the simulation accurately models diffusion in the pore space of the experiment. We show the results obtained with this procedure in Fig.~\ref{from sep}, for two typical experiments. This figure shows experimental (solid lines) and simulated (dashed lines) drying rates, where the latter are extracted by taking the experimental pattern sequence recorded in $T_{ij}$ and using it to estimate the drying rate with our model, based on the resulting vapor concentration gradients. This process clearly improves our estimate, getting experimental and simulated drying rates to agree. The improvement suggests that diffusion is actually well modeled in our simulations, allowing us rule out one possible source of error and focus on invasion mechanisms and the formation of isolated clusters as sources of the observed discrepancy in drying rates. 

\subsection{Wettability}

A potential cause for the more limited appearance and the lack of persistence of isolated liquid clusters in simulations are wettability effects that are not considered in the model. To examine this, we repeated some experiments with a less wetting fluid, water, which has a contact angle of about 70\textdegree \ on NOA \cite{DavideNOA} (as opposed to the $\simeq 3$\textdegree \ contact angle of 3M Novec 7500). A highly wetting fluid, like our oil, easily breaks into clusters, especially in the corners between pillars and the walls of our cells, forming rings of fluid around the bases of the pillars. These rings may not influence fluid transport directly, but they could contribute by increasing vapor concentration in pores, thus affecting drying behavior. When the wettability of the fluid decreases (e.g when using a different liquid) the formation of these rings and isolated clusters decreases as well, as a consequence of the higher surface tension of the liquid.  

Results are shown in Fig.~\ref{repeat+water}, which shows the metrics for one particular experiment run three times with oil (blue cloud) and once with water (black line). It can be seen that the fluid saturations, the area-to-perimeter ratios of the leading fronts, and the drying rates behave virtually the same way in both cases. The only observed difference is in the Euler number. Not only is the final amount of clusters lower when using water, but the number of clusters that are formed is always lower at the same saturation. This shows that a higher liquid surface tension alone can limit cluster formation.

\section{Discussion and conclusions}

The goal of this paper was to introduce a novel experimental method to study drying in porous materials. We have used microfluidic cells to make a two-dimensional granular medium with grains of size comparable to those of real soils, improving on both the control of the sample geometry of previous works \cite{expglassbeads, Shaw}, the pore sizes \cite{Ferrariexp,aursjo2011direct} the throat sizes \cite{ExpVorhauer,Laurindo1998,vorhauer2015drying} and the number of objects in the porous medium \cite{12experiment}. The improvement was made possible by using microfabrication techniques common in microfluidics \cite{Madou} that allow precise control over the positioning of micron-scale objects. Using these methods we have also imposed a heterogeneity on the size distribution of the grains and pores, in order to investigate the effects of disorder on drying behavior in porous media. The microfluidic cells were filled with a volatile oil and allowed to dry. The experiments were then compared to pore-network model simulations, of the same geometries. This class of models was chosen thanks to its ability to combine computational efficiency and description of complex effects~ \cite{nowicki1992microscopic,Prat1993}.

The order of magnitude of the initial drying rates, when the porous nature of the system is not as important in the evaporation process \cite{Shahraeeni}, is captured well by the PNM (Fig. \ref{e0 vs BL}). We attribute the variability of experimental initial drying rates to distortions of the boundary layer's thickness that occur during the manufacturing process, whereas a more systematic discrepancy between experimental and simulated initial drying rates can be reduced by considering a small, additional boundary layer extending outside the manufactured cell. This discrepancy can be accounted for by including it in a scaling of the drying rates by the initial evaporation rate. 

The experiments were also able to capture what is sometimes argued to be a three-dimensional effect -- the distinction between the constant and falling rate periods \cite{Shokri2010a,Chauvet2009,shokri2011determines} -- that two dimensional numerical models typically are not able to reproduce, in particular in the absence of film flow \cite{Prat2007}. The experimental drying rates shown in this paper (Fig. \ref{sat_rate disorder}) are often constant throughout the duration of experiments, although we did observe several instances where the onset of a falling rate period was visible. The formation and persistence of isolated liquid clusters within the dry area alters the vapor concentration in the porous medium, sustaining a faster drying rate than that predicted by the PNM. On the other hand, simulations do not reproduce this trend, slowing down their drying rate early as evaporation occurs at the pore-space surface, until the loss of surface wetness. After this point, vapor diffusion within the medium is the dominant mechanism by which fluid is removed \cite{Chauvet2009,shokri2011determines}.  Furthermore, isolated liquid clusters tend to evaporate at a higher rate in simulations than in experiments (Fig. \ref{MM}b). This causes the drying front to effectively recede deeper into the medium, causing the drying rates to drop again due to the relative lower efficiency of vapor diffusion as a fluid-removing mechanism.

Since the evolution of the evaporation rates were generally not reproduced well in the PNM, we have also investigated the agreement between the experimental and simulated drying \textit{patterns}. We have used the Minkowski measures to compare them, as metrics suitable to characterize complex binary patterns \cite{MeckeMinkowski,RalfKarinMinkowski}. Their statistical descriptions of the leading patterns were reproduced very well (Fig. \ref{MM}), and we have further quantified this match by comparing the patterns pore-by-pore (Fig. \ref{pore-pore}) at the same main cluster saturation, $S_0$. We have shown how, given the manufacturing tolerance in our experiments, the highest expected match in the leading patterns is 70-80\%. We do, in fact, reach such a match in most patterns at breakthrough.  By looking at the effects of random errors, we have suggested that in the cases where this threshold is not reached, point-errors in throat sizes can be responsible for the different evolution of the drying front into the porous medium; the patterns themselves are highly sensitive to disorder.

One noteworthy discrepancy between our experiments and simulations is the number of isolated liquid clusters in the dry area. We have shown how the persistence of these clusters in the experiments influences drying rates (e.g. Fig. \ref{from sep}). We have also investigated various origins of these clusters. In particular, we tested the possibility of wettability effects, which are not included in the model, on drying by repeating experiments with cells filled with water instead of oil.  This changed the contact angle of the liquid from $\sim$0\textdegree \ to $\sim$70\textdegree. The resulting drying rates and leading patterns were indistinguishable from the same experiment with oil. The only noticeable difference was in the number of clusters formed: water did not break down into as many isolated clusters, due to its higher surface tension.  We also saw how these isolated clusters still tended to evaporate slower than the model.  This is demonstrated in Fig.~\ref{repeat+water}, in which the Euler number drops throughout drying, as opposed to the trend shown by simulations in Fig.~\ref{MM}, where the Euler number consistently increases in the later stages of drying. 

A higher contact angle inhibits the formation of fluid films and gutter flows, which may otherwise enhance liquid transport within the porous medium \cite{Prat2007}. In the wettability range explored, inhibiting this mechanism only changed the number of clusters formed, but not the evaporation rate or leading pattern of the experiment.  This shows that if gutter and film flows were present in the experiments with oil, then they would only play a negligible role in the fluid transport. Besides this, the observation of a comparable relative drying rate, despite the different numbers of isolated clusters, suggests that the positions, rather than the density, of clusters predominantly influences the drying rate. This statement is confirmed by the data in Fig.~\ref{from sep}. There, the drying rates were computed by taking an experimental pattern and estimating the drying rate at the next time step. This procedure allows for a much closer estimation of experimental rates by the model, proving the important effect of the pattern of isolated clusters on the drying rate.  

Finally, we investigated the effects of adding random heterogeneities to drying porous media. Random disorder does not show any clear effect on the metrics that we have used to characterize our experiments: there is no faster drying rates for higher disorder, for example. Neither did simulations show any noteworthy trends, even though they allow for averaging over many realizations.  However, in a companion paper we investigate the effects of \textit{correlated} disorder, on drying  \cite{Borgman2016}.  Using similar experiments and methods to those reported here, there we demonstrate that if there are local patches of larger-than-average or smaller-than-average pores, grouped together over some correlation length, then this additional length-scale of the the heterogeneity in a porous medium can significantly affect drying rates, and drying patterns.

Even though 2D experiments have been carried out \cite{Ferrariexp, ExpVorhauer, expglassbeads, Shaw} for similar phenomena before, this is, to our knowledge, the first time that large two-dimensional experiments, with length scales comparable to those of soils have been performed and directly coupled with PNM. We were able to these produce microfluidic chips with high precision, making such comparisons simpler thanks to the ability of reproducing the same designs in the simulated samples.

These methods can be quite generally applied, and we expect that microfluidic techniques can be used to study the broad class of problems where experimental micro-models have traditionally been applied.  This is obviously not limited to drying, but includes fluid-fluid displacement or salt transport and deposition.  The techniques offer improvements on existing experiments \cite{holtzman2015wettability,zhao2016wettability} by allowing one to reach length scales comparable to those of soils, where most of the applications of these problems are. Such two-dimensional microfluidic micromodels also allow for direct validation of the types of numerical models that are widely used in studying granular packings, saving the time necessary to manufacture the samples and perform simulations by simplifying analysis, and allowing for the removal of unnecessary complications and uncertainties.

\acknowledgments
This joint research project was financially supported by the State of Lower Saxony, Hannover, Germany (\#VWZN2823).  RH also acknowledges partial support from the Israeli Science Foundation (\#ISF-867/13) and the Israel Ministry of Agriculture and Rural Development (\#821-0137-13).


\end{document}